\documentclass[prl,aps,twocolumn,preprintnumbers,amsmath,amssymb]{revtex4}
\usepackage{graphicx,psfrag}
\begin{document}
\preprint{UK/TP 06-18}

\title{\boldmath%
Observing Dark Matter via the Gyromagnetic Faraday Effect
\unboldmath}

\author{Susan Gardner}

\affiliation{Department of Physics and Astronomy, University of Kentucky, 
Lexington, KY 40506-0055 
}


\begin{abstract}
If dark matter consists of
cold, neutral particles with a non-zero magnetic moment, then, 
in the presence of 
an external magnetic field, 
a measurable gyromagnetic Faraday effect becomes possible. 
This enables direct constraints on 
the nature and distribution of such 
dark matter 
through detailed measurements of the polarization and temperature 
of the cosmic microwave background
radiation. 
\end{abstract}

\maketitle

{\bf{\em Introduction.~}}
The existence of dark matter (DM) was first inferred 
in 1933 from Zwicky's observations of extragalactic 
nebulae~\cite{Zwicky:1933gu}. 
In recent years, our ability to assay its abundance has
sharpened considerably, and 
a concordance of disparate observations reveal that 
DM 
comprises some 
twenty-three percent 
of the energy density of the universe,
with a precision of a few percent~\cite{concord}. 
Yet, despite this progress, 
the fundamental nature of 
DM remains unclear. 
One cannot say whether DM
consists of a single species of particle, 
or of many, or even if 
it consists of stable, elementary particles 
at all. Dark matter could comprise aggregates of some kind, or 
be mimicked, in part, 
by a modification of gravity at large distances~\cite{modgrav,arkhad,bullet}. 
We do know that light, massive neutrinos 
cannot explain the galactic rotation curves~\cite{Tremaine:1979we}, 
so that 
non-Standard-Model particles, arguably of the Fermi scale, 
are commonly invoked to explain it~\cite{Feng:2005uu}. 
Accordingly, little, if anything, 
is known of each species' quantum numbers, 
mass, or mass distribution. 
In this Letter  
we consider the possibility that 
DM consists 
of neutral 
objects, which need not be elementary particles, 
of mass $M$ with non-zero magnetic moments. 
The empirical limits on this possibility 
vary with the particle's mass~\cite{Pospelov:2000bq,Sigurdson:2004zp}
and can be evaded if the particle is composite. 

Although our scenario naturally permits the dark constituents to be 
mutually interacting~\cite{Spergel:1999mh}, 
it does differ significantly from usual ideas. 
For example, models 
of electroweak symmetry breaking with an additional discrete
symmetry can yield viable 
DM candidates. 
In models with supersymmetry, 
the DM
candidate --- the ``lightest supersymmetric particle'' --- 
is a Majorana particle, and its static magnetic
moment is identically zero. 
Thus if the effect we discuss is observed, 
it demonstrates that supersymmetry does not provide 
an exclusive solution to the 
DM problem. 
On the other hand, models with ``large'' extra dimensions, 
such that their compactification radius $R$ has $R^{-1} \lesssim 1$ TeV, 
offer DM candidates 
which are nominally consistent with our 
scenario~\cite{Arkani-Hamed:1998vp}. 
In particular, models with universal extra dimensions~\cite{Appelquist:2000nn} 
yield DM 
candidates which are known to be compatible with
observed constraints and which could also possess 
magnetic moments~\cite{Servant:2002aq,Cheng:2002ej,Hsieh:2006qe}. 

Let us now consider how cold 
DM with a non-zero magnetic
moment can be observed. 
A medium of particles with 
either electric charges or 
magnetic moments develops a circular birefringence when subjected
to an external magnetic field, even if the medium is isotropic. 
Consequently, the 
propagation speed of light in the medium will depend
on the state of its circular polarization, so that 
light prepared in a state
of linear polarization will suffer a rotation of
the plane of that polarization upon transmission through
the medium. If we define $k_\pm$ to be the wave number 
for states with right- ($+$) or left-handed ($-$) circular 
polarization, then
the rotation angle is given by $\phi = (k_+ - k_-)l/2$, 
where $l$ is the length of transmission through the medium. 
If the medium contains free electric charges, this is the 
Faraday effect known for light
travelling through the electrons and magnetic fields 
of the warm interstellar medium (ISM)~\cite{firstFR}. 
A Faraday effect can also
occur in a magnetizable medium which is electrically 
neutral~\cite{polder,hogan}.
We term these the gyroelectric (GE) and gyromagnetic (GM) Faraday 
effects~\cite{EEbook}, respectively. We study the GM 
Faraday effect associated with cold 
DM carrying a non-zero
magnetic moment.
We begin by comparing the
Faraday effects in the ISM, for which the GE
effect
is familiar, 
before turning to a discussion of their impact
on the cosmic-microwave background (CMB) polarization
and the constraints such measurements can 
yield on models of DM.

{\bf{\em Faraday Effects in the ISM.~}}
The ISM contains free electrons and external 
magnetic fields; it is GE 
and gives rise to a Faraday effect. 
We consider an external magnetic field $\mathbf{H}_0$ 
in the $\hat{\mathbf{x}}$-direction with 
circularly polarized electromagnetic
waves propagating parallel to it. In this case, 
an electron with charge $-e$ and mass $m$ suffers a 
displacement $\mathbf{s}$ via the Lorentz force 
\begin{equation}
m \ddot{\mathbf{s}} = -e (\mathbf{E} + \dot{\mathbf{s}}\times
 \mathbf{H}_{\rm tot}) \,,
\label{lorentz}
\end{equation}
where 
$\mathbf{H}_{\rm tot} = \mathbf{H}_0 + \mathbf{H}$. 
The electric field, e.g., associated with the wave 
is 
$\mathbf{E}(\mathbf{x},t) = E_\pm \mathbf{e}_{\pm} 
\exp(i k_\pm x -i \omega t)$, 
where $\mathbf{e}_\pm \equiv \hat{\mathbf{y}} \pm i \hat{\mathbf{z}}$. 
We define the polarization state with positive helicity, $\mathbf{e}_+$, 
to be right-handed.
Assuming 
$|\mathbf{H}_0|\gg |\mathbf{H}|$, the steady-state solution for 
$\mathbf{s}$ yields, for a medium of electrons with
number density $n_e$, 
 the polarization $\mathbf{P} = -n_e e \mathbf{s}$ 
and the electric susceptibility $\chi_e$, recalling 
$\mathbf{P}_\pm = \epsilon_0 \chi_{e\,\pm} \mathbf{E}_\pm$. 
We thus determine the permittivity $\epsilon_\pm$: 
\begin{equation}
\frac{\epsilon_\pm}{\epsilon_0} 
\equiv 1 + \chi_{e\,\pm} = 1 
- \frac{\omega_P^2}{\omega (\omega \mp \omega_H)} \,,
\end{equation}
where the plasma frequency $\omega_P$ is given by 
 $\omega_P^2 \equiv n_e e^2/\epsilon_0 m$
and $\omega_H = e H_0/m$. With 
$k_\pm = (\omega/c) \sqrt{\epsilon_\pm/\epsilon_0}$ 
and with $\omega \gg \omega_H, \omega_P$, 
we have $\phi = -{\omega_P^2 \omega_H l}/{2c\omega^2}$ 
to leading 
order in $\omega$. 
Generalizing this to variable electron densities
and magnetic fields along the line of sight yields 
\begin{equation}
\phi = -\frac{e^3}{2c \omega^2 \epsilon_0 m^2} 
\int_{0}^{l} dx\, n_e(x) H_0 (x)  \,,
\end{equation} 
where $x=0$ marks the location of the source. 
The $\omega$ dependence makes knowledge of the
intrinsic source polarization unnecessary; one measures 
the position angle of linear polarization, in a fixed 
reference frame, as a function of $\omega$, so that the 
line integral of $n_e(x) H_0(x)$ can be 
inferred~\cite{FR,Kronberg:1993vk}. 
A pulsed radio source also permits the measurement of the
frequency dependence of the arrival time, 
to yield the 
line integral of $n_e(x)$~\cite{FR}, so that the average magnetic field
along the line of sight can also be determined. 

If the electrons can be aligned to yield a magnetization, 
the ISM 
can be regarded as 
GM 
as well. We shall treat 
the GE and GM 
effects independently. 
Applying a magnetic field in a GM 
medium
induces a magnetization $\mathbf{M}_{\rm tot}$, i.e., 
a net magnetic moment/volume, where 
$\mathbf{M}_{\rm tot} = \hat{\mathbf{x}} M_0 + \mathbf{M}$ 
and $M_0$ results from $H_0$ alone. 
The resulting magnetization
obeys 
\begin{equation}
\dot{\mathbf{M}}_{\rm tot} = \gamma 
\mathbf{M}_{\rm tot} \times \mathbf{H}_{\rm tot} \,,
\label{larmor}
\end{equation}
where $\gamma$ is the 
gyromagnetic 
ratio of the magnetic-moment-carrying
particle. 
If the 
constituents possess an electric dipole moment as well,
an additional term appears in Eq.~(\ref{larmor})~\cite{Bargmann:1959gz}.
We assume $|\mathbf{H}_0|\gg |\mathbf{H}|$, 
$|\mathbf{M}_0|\gg |\mathbf{M}|$, 
and the conventions of the GE 
case 
to determine the steady-state solution, which, 
neglecting the 
$\mathbf{M} \times \mathbf{H}$ term, is 
\begin{equation}
M_\pm = \pm  \frac{\chi_0 \omega_H}{\omega \pm \omega_H} H_\pm 
\equiv \chi_\pm H_\pm \,,
\end{equation}
where $\chi_0 \equiv M_0/H_0$ and 
$\omega_H \equiv \gamma H_0$. We recall the magnetic susceptibility
$\chi_m$ obeys $\mathbf{M} = \chi_m \mathbf{H}$, so that 
\begin{equation} 
\frac{\mu_\pm}{\mu_0} \equiv 1 + \chi_{m\,\pm} = 1 \pm 
\frac{\chi_0\omega_H}{\omega \pm \omega_H} \,,
\end{equation}
where $k_\pm =(\omega/c)\sqrt{\mu_\pm/\mu_0}$. 
Noting $\omega_H/\omega \ll 1$ and 
working to leading order in 
this quantity, 
one has $k_{\rm diff} = k_+ - k_-$, which controls 
$\phi$,  with 
\begin{equation}
k_{\rm diff} 
= \frac{\chi_0 \omega_H}{c}  +
\frac{\chi_0 \omega_H^3}{c\,\omega^2} + 
\frac{\chi_0^2 \omega_H^3}{2c\,\omega^2} + \dots \,.
\label{magkdiff}
\end{equation}
The magnetization induced by $H_0$ on a system of spin-$1/2$ particles each
with magnetic moment $\mu$
in equilibrium at temperature $T$ is~\cite{polreview}
\begin{equation}
M_0 = n_e \mu \tanh \left( \frac{\mu H_0}{k_B T}\right) = 
n_e \left( \frac{\mu^2 H_0}{k_B T}\right) \,,
\label{magnet}
\end{equation} 
where the corrections to the last equality are negligible 
in the ISM, though diverse environmental conditions do exist. 
The magnetic field $H_0$ is no larger than a few $\mu$G
--- and its cold patches are no colder than a few 100 $K$~\cite{FR}. 
We can thus neglect non-leading powers in $\chi_0$.
We separate the rotation angle $\phi$ into 
frequency-independent and frequency-dependent pieces, so that
$\phi = \phi_0 + \phi_\omega$, to yield 
\begin{eqnarray}
\phi_0 &=& 
\frac{\mu^2 \gamma}{2c k_B} \int_0^l dx 
\frac{n_e (x) H_0(x)}{T(x)} \,,
\label{phi0}
\\
\phi_\omega &=& 
\frac{\mu^2 \gamma^3 }{2c \omega^2 k_B} \int_0^l dx 
\frac{n_e (x) H_0^3(x)}{T(x)} \,, 
\label{phiomega}
\end{eqnarray} 
where $\gamma = g \mu_B/\hbar$, $\mu_B\equiv e/2m$, and $g$ is the usual 
Land\'e factor.  
The appearance of higher powers in
$H_0$ in $\phi_\omega$ makes it, as well as the time delay, negligible 
in comparison to $\phi_0$ in the ISM. 
If we neglect any 
$T$ variation along the line of sight, then 
the frequency-independent GM and GE
effects share a common integral.
We can then compare 
\begin{equation}
\phi_0 = 
\frac{\mu^2 \gamma}{2c k_B T} \int_0^l dx \,
n_e (x) H_0(x)
\end{equation} 
with 
\begin{equation}
\phi = \frac{e^3}{2c \omega^2 \epsilon_0 m^2} 
\int_{0}^{l} dx\, n_e(x) H_0 (x)  \,
\end{equation} 
by computing 
\begin{equation}
|\tilde{\chi}| \equiv 
\frac{\gamma \mu^2}{k_B T}  = 
\frac{|g| \mu^2 \mu_B}{\hbar k_B T} \sim \frac{2 \mu_B^3}{\hbar k_B T} \sim 
4.6 \cdot 10^{-19} 
\left[\frac{300\, \hbox{K}}{T}\right]
\frac{\hbox{cm}^3}{ \hbox{G s}} 
\nonumber
\end{equation} 
and 
\begin{equation}
{\chi} \equiv
\frac{e^3}{\omega^2 \epsilon_0 m^2}
= 
\frac{\alpha}{\pi} \cdot 
\frac{\hbar }{m c}\cdot \frac{e}{m } \cdot \lambda^2 
\sim 1.6 \cdot 10^{-6} \left[\frac{\lambda}{1\,\hbox{cm}}\right]^2 
\frac{\hbox{cm}^3}{ \hbox{G s}} \,, 
\nonumber 
\end{equation}
where $|g|\sim 2$, $\mu_B \sim 5.79\cdot 10^{-9}$ eV/G, 
$k_B T \sim 1/38.7$ eV for $T\sim 300$ K, 
$1\, \hbox{eV} \sim 4.03 \cdot 10^{-11}\, 
\hbox{G}^2 \hbox{cm}^3$, $\alpha \sim 1/137$, 
$e/m \sim 1.76 \cdot 10^7$ rad/Gs, 
and $\hbar/mc \sim 3.86 \cdot 10^{-11}$ cm. 
Recent surveys have used wavelengths in the 
$\lambda =6$ and $20$ cm bands~\cite{FR,Clarke:2000bz}, and 
most Faraday rotation
accrues in the warm ISM, for which $T \sim 5000$ K. We thus
find the GM 
effect to be 
negligible for radio sources. 
We note $\phi_\omega$ is smaller than $\phi_0$ by a factor of 
$\gamma^2 H_0^2/\omega^2 
\sim 9 \cdot 10^{-21} [\lambda/(1\,\hbox{cm})]^2$, using 
$H_0 \sim 10^{-6}$ G.

{\bf{\em Faraday Effects on the CMB Polarization.~}}
Our study of the GM Faraday effect shows 
$\phi_0$ to be the 
most important numerically, though its frequency independence means 
we must employ sources of known polarization to determine it. 
To realize this, we turn to the CMB radiation, for 
the scalar gravitational perturbations 
which dominate 
the temperature fluctuations in inflationary cosmologies 
give rise to $E$-mode, or gradient-type, 
polarization exclusively~\cite{seljak,kamio}. 
The Faraday effects provide a mechanism by which $B$-mode, or curl-type, 
polarization 
can be produced from an initial state of $E$-mode polarization; 
ultimately, we wish to 
interpret the $B$-mode polarization as a constraint on 
DM with a 
magnetic 
moment. 
A variety of sources of $B$-mode polarization exist, however, and 
it is important to separate the possibilities. 
Let us enumerate some of them explicitly. 
Primordial tensor 
or vector 
gravitational perturbations  
in the CMB can give rise to $B$-mode polarization~\cite{seljak,kamio},
and $B$-mode polarization can be generated from primordial 
$E$-mode polarization via gravitational lensing~\cite{zalda}.  
Magnetic fields can also imprint $B$-mode polarization. 
Primordial magnetic fields can do this both through the perturbations
they engender~\cite{primo}, 
as well as through the GE 
Faraday rotation they mediate~\cite{primoFR}. 
Magnetic fields in galactic clusters~\cite{Clarke:2000bz} 
can also give rise to GE 
Faraday rotation~\cite{Takada:2001bm}, impacting the 
CMB polarization at small angular scales~\cite{seljak,kamio,huwhite}. 
The GE 
Faraday effect is distinguished by
its $\omega^{-2}$ frequency dependence; the $B$-polarizations
engendered by gravitational lensing and radiation are frequency-independent. 

The GM 
Faraday effect 
can operate if the medium 
has 
a magnetization; this can occur if a non-zero magnetic field exists
while the DM is still in thermal equilibrium in the early Universe. 
These conditions suffice to polarize it to the degree
given by Eq.~(\ref{magnet}); for cold DM, $H/T$ is a constant
over the cosmological expansion. Once
DM decouples there is no mechanism to polarize it, and 
its primordial polarization cannot be lost. The primordial magnetic field
changes slowly with respect to the Larmor precession rate, so that the
DM magnetization
can track the magnetic field as the Universe evolves. In constrast, the 
electron's charge drives $e + p \leftrightarrow H +\gamma$ as the
Universe cools and washes out any primordial polarization it possesses.
At much later time scales, the reionized electrons may acquire a non-zero
magnetization, but their dilute nature 
make the associated B-mode polarization
immeasureably small. 
Thus a non-zero, frequency-independent B-mode polarization, induced
through GM Faraday rotation, 
can be 
attributed to the presence of DM with a non-zero magnetic 
moment. This effect, in turn, is signalled by
the presence of frequency-independent 
$EB$ cross-correlation power spectra in the CMB. 
Recent studies of WMAP
and BOOMERANG data provide mild evidence for 
this effect~\cite{bfeng}, and future studies 
at PLANCK and CMBpol
can provide sharpened constraints~\cite{planck}. 

{\bf{\em Constraining DM.~}} 
We now consider how the GM 
Faraday effect can be  
used to constrain models of 
DM. To evaluate the Faraday 
rotation we must integrate over the past light cone of the photon, 
including the cosmological scale dependence of the DM density. 
We consider the Faraday rotation accrued through the transit of 
cold DM, 
so that the scale dependence of the magnetic field
and temperature cancel; we 
assume, moreover, the ratio of these quantities
to be constant. 
Thus we modify 
Eq.~(\ref{phi0}) 
\begin{equation} 
\int_0^l dx n(x) \rightarrow n_o 
c \int_0^z\, 
{dz^\prime}{{H}(z^\prime)}^{-1} (1 + z^\prime)^3
\equiv n_o \tilde l \,
\end{equation}
to define the effective path length $\tilde l$, so that 
$\phi_0 = \mu^2 \gamma {H}_{\rm prim}^o n_o \tilde l / 2c k_B T^o $, 
where ${H}_{\rm prim}^o$, $n_o$, and $T_o$ are 
the primordial magnetic field, DM number density, and temperature, all 
scaled to the present epoch. 
We solve for $H(z)$, the Hubble constant at a redshift of $z$, 
using the Friedmann equation in a flat $\Lambda$CDM cosmology 
with a matter energy density of $\Omega_M=0.27$
and with $H_o= 71\,\hbox{km}\,\hbox{s}^{-1}\hbox{Mpc}^{-1}$~\cite{wmap}. 
For a spin-$1/2$ 
DM particle
of mass $M$ we define 
the magnetic moment $\mu \equiv\kappa\mu_M $, 
so that $\kappa$ is the Pauli moment, as well as 
the gyromagnetic ratio $\gamma = 2\kappa \mu_M /\hbar$ with $\mu_M = e/2M$. 
DM has been established in the recombination era~\cite{concord}, 
so that we compute 
the angle $\phi_0$ engendered by CMB photons propagating from 
$z\sim 1100$ to the present. To estimate the present-day DM temperature 
we consider galactic DM and use 
the gravitational infall velocity, 
assuming a Maxwell-Boltzmann distribution in galactic DM velocities, 
to determine 
that the root-mean-square velocity 
obeys $v_{\rm rms}/c = \sqrt{3 k_B T/Mc^2}$. 
Thus 
\begin{eqnarray}
\phi_0 &\sim& 3.6\cdot 10^{-18}\, 
\frac{\hbox{cm}^3}{\mu\hbox{G}\, \hbox{Mpc}}
\left(\frac{\mu}{\mu_B}\right)^3 \left(\frac{m}{M}\right)^2  
\left(\frac{v_{\rm rms}}{c}\right)^{-2}  
\nonumber \\
&\times& n_o [\hbox{cm}^{-3}] { H}_{\rm prim}^o [\mu\hbox{G}] 
\tilde l [\hbox{Mpc}]  \,, 
\end{eqnarray}
where 
$ n_o \sim 2.17 \cdot 10^{-3} \,\hbox{cm}^{-3}$, noting 
$ n_o \equiv \rho_{\rm{cdm}}/m_e$ and 
$\rho_{\rm{cdm}} \sim 1.98 \cdot 10^{-30} \hbox{g}\,\hbox{cm}^{-3}$~\cite{wmap}, 
$v_{\rm rms}\sim 200$ km/s, and 
$\tilde l \sim 1.3 \cdot 10^{10}$ Mpc. 
We can 
consider 
light cold dark matter because 
its annihilation cross section 
is mediated by its magnetic moment~\cite{leew}. 
Some observational 
evidence suggests that $M$ is of MeV scale~\cite{Beacom:2004pe}. 
Using 
the bound ${H}_{\rm prim}^o \lesssim 10^{-3}\,  \mu\hbox{G}$, 
for primordial magnetic fields 
coherent across the present horizon~\cite{Blasi:1999hu}, 
we find a bound of 
$|\kappa| \lesssim 0.8$  if $M=m/10$ and 
if $\phi_0$ can be determined to $\phi_0 \sim 10^{-2}$ rad. 
Precision electroweak measurements also constrain the
magnetic moment~\cite{Sigurdson:2004zp}. 
The quantity $\Delta \hat r$ represents the radiative corrections 
to the relationship between the fine-structure constant 
$\alpha$, the Fermi constant $G_F$, and the $W^\pm$ 
and $Z$ masses, $M_W$ and $M_Z$~\cite{Marciano:1999ih}; 
the difference 
between the empirically determined value of $\Delta \hat r$
and that computed in the Standard Model 
provides a window $\Delta \hat r^{\rm new}$ to which a DM
particle can contribute. 
Thus we find 
from the vacuum polarization 
correction to the photon self-energy, with $a\equiv (M_Z/M)^2 \gg 1$,  
\begin{equation}
\Delta \hat r^{\rm DM} \sim - \kappa^2 \frac{\alpha}{4\pi} 
\left( \frac{a}{6}\log a - \frac{a}{9} + O(1)\right)
\left(1  - \frac{M_Z^2}{M_c^2}\right)^{-4} \nonumber\,,
\end{equation}
where we include a form factor at each vertex with a compositeness
scale of $M_c$.
With $\Delta \hat r^{\rm new} < 0.0010$ at 95\% CL~\cite{pdg2006},  
we find with $M=m/10$ that 
$|\kappa| < 3.4\cdot 10^{-7}$ if $M_c \to \infty$,
which relaxes to 
$|\kappa| < 1.5$ if 
$M_c = 2$ GeV, e.g. 
We thus conclude that
a useful constraint on $\kappa$ 
from a 
Faraday rotation measurement is possible. 

{\bf{\em Summary.~}}
A Faraday effect also exists for light transiting a dark medium of 
electrically neutral particles with non-zero magnetic moments in an external
magnetic field. We have shown that 
this possibility can 
serve as a new source of $B$-mode polarization in the CMB
and that it can be 
disentangled from 
other sources. Thus 
a non-zero effect due to such 
DM can be identified, if it exists, 
with the implication that supersymmetric models do not
provide an exclusive solution to the 
DM problem. 
The GM 
Faraday effect can be used to probe
the nature and distribution of DM, 
to realize a picture of our Universe 
shaped by what we {\it observe}, rather than by 
what we believe to be so.

{\bf{\em Acknowledgments.~}} 
S.G. is grateful to 
T. Rizzo, T. Troland, 
S. Brodsky, S. Church, J. Hewett, J. Nico, K. Olive, 
H. Quinn, and A. Zee 
for encouragement and useful comments, to 
J. Hewett and S. Church for reading the manuscript,
and to G. Jones and Z.-T. Lu for discussions of 
polarization techniques. 
She thanks the SLAC theory group and the Institute for
Nuclear Theory for gracious 
hospitality and acknowledges partial support from the 
U.S. Department of Energy under 
contract DE--FG02--96ER40989.

\end{document}